\documentclass[aps,pra,twocolumn,showpacs,preprintnumbers,amsmath,amssymb,superscriptaddress,longbibliography]{revtex4-2}
\usepackage{mathrsfs}
\usepackage{amsfonts}
\usepackage{amsmath}
\usepackage{amssymb}
\usepackage{natbib}
\usepackage{bbm}
\usepackage[normalem]{ulem}
\usepackage{hyperref}
\usepackage{xcolor}
\usepackage{graphicx}

\begin{document}
	\title{Neutral atom entangling gate in the ultrastrong coupling regime}

	
	\author{Ebubechukwu O. Ilo-Okeke} \email{ebube@wm.edu}
	\affiliation{Department of Physics, College of William and Mary, Williamsburg, Virginia 23187, USA.}  
	\affiliation{New York University Shanghai, NYU-ECNU Institute of Physics at NYU Shanghai, 567 West Yangsi Road, Shanghai, 200124, China.}

	\author{Tongzhou Wang}
	\affiliation{New York University Shanghai, NYU-ECNU Institute of Physics at NYU Shanghai, 567 West Yangsi Road, Shanghai, 200124, China.}

    \author{Valentin Ivannikov}
	\affiliation{New York University Shanghai, NYU-ECNU Institute of Physics at NYU Shanghai, 567 West Yangsi Road, Shanghai, 200124, China.}
	
	\author{Tim Byrnes}\email{tim.byrnes@nyu.edu}
	\affiliation{New York University Shanghai, NYU-ECNU Institute of Physics at NYU Shanghai, 567 West Yangsi Road, Shanghai, 200124, China.}
	\affiliation{State Key Laboratory of Precision Spectroscopy, School of Physical and Material Sciences, East China Normal University, Shanghai 200062, China}
	\affiliation{Center for Quantum and Topological Systems (CQTS), NYUAD Research Institute, New York University Abu Dhabi, UAE.}
	\affiliation{Department of Physics, New York University, New York, NY 10003, USA}

 	\date{\today}

	\begin{abstract}	
	We propose a method to deterministically entangle qubits or ensembles of qubits interacting with a shared bosonic mode in the ultrastrong coupling regime. We show that the resulting gate is a product of two unitaries: one unitary acts only on the quantum state of the qubits and entangles them, while the other acts only on the quantum state of the boson, producing a phase shift. We find that the gate time is inversely proportional to the qubit-boson interaction strength, and by tuning the qubit-boson interaction strength, one can prepare a maximally entangled state or a squeezed state. Applying the quantum gate to multiple qubit ensembles, we show that the quantum gate prepares a Schr\"odinger cat state. We also examine imperfections such as including free evolution of the qubits, and show that this produces an effective mixing.   Our proposal is feasible for ultrastrong coupling experiments. 
	\end{abstract}

    \maketitle
	\section{Introduction}\label{sec:Introduction}
	Quantum gates are the building blocks of quantum computing~\cite{nielsen2000,williams2011}, analogous to classical logic gates in traditional computers. These gates perform unitary evolution on qubits, exploiting the principles of superposition and entanglement to process information in ways that classical systems cannot. Various quantum gates~\cite{barenco1995} have been developed, each with distinct functionalities that enable complex quantum algorithms. The basic types of quantum gates include single-qubit gates, which manipulate the state of an individual qubit, and multi-qubit gates or entangling gates, such as the CNOT, which enable interactions between qubits.

	The development and physical implementation of quantum entangling gates has followed different routes on various platforms~\cite{cirac1995,sorensen1999,saffman2010,cirac1995,briegel2009,haroche2003,blais2004,byrnes2015}. For example, in trapped ion computing~\cite{cirac1995,sorensen1999,sorensen2000}, the M{\o}lmer-S{\o}rensen gate~\cite{sorensen1999,sorensen2000} is a widely used entangling quantum gate. It is realized by applying bichromatic laser fields to the ions, which induce a spin-dependent force that couples the internal states of the ions with their vibrational modes. This force enables controlled interactions between trapped-ion qubits that lead to entanglement. Other gates~\cite{garcia-ripoll2003,garcia-ripoll2005,hussain2014,hussain2015} in trapped ions and neutral atom computing use geometric phases, where two or more trapped ions or atoms entangle to light. In the technique,  the light mode at the end of gate operation is the same as that at the beginning up to a global phase. However, the quantum states of the atoms or ions become entangled with each other, while disentangled from the light state. Similarly, entangling gates in neutral atom computing in optical tweezer arrays~\cite{kaufman2021,bluvstein2024} are realized by exciting an electron to a high principal quantum number, leading to strong dipole-dipole interactions between atoms~\cite{jaksch2000,lukin2001}. These interactions enable the implementation of entangling quantum gates through the Rydberg blockade~\cite{wilk2010,isenhower2010}. Other proposals using quantum measurements~\cite{brune1992,aristizabal-zuluga2021,kuzmich2000,blais2004,byrnes2015} involve a nearly resonant two-level system, where the coupling strength of the matter-oscillator is small compared to an oscillator and qubit frequencies uses an ac Stark shift~\cite{kuzmich1998,blais2004,ilo-okeke2014,ilo-okeke2024} to produce entanglement between multiple atomic systems or manipulating light states~\cite{davidovich1994,julsgaard2001,ilo-okeke2021,ilo-okeke2022}.
	
	The ultrastrong coupling regime~\cite{forn-diaz2019,kockum2019} is an emerging platform on trapped ions~\cite{lv2018}, superconducting qubits~\cite{yoshihara2017}, and neutral atoms~\cite{baumann2010,koch2023,hunanyan2024}, showing promise for quantum computation. Here, the interaction strength between matter and the bosonic field becomes comparable to or exceeds the bosonic field frequency and the transition frequencies of the qubits. This regime goes beyond the Jaynes-Cummings model~\cite{shore1993}, which neglects the counter-rotating terms in the interaction Hamiltonian. Several studies~\cite{yoshihara2017,lv2018,koch2023,hunanyan2024} have shown that the inclusion of the counter-rotating terms produces novel physical phenomena, such as generation and bouncing of phonon wave packets~\cite{lv2018}, quantum phase transitions~\cite{baumann2010,baumann2011}, the Bloch–Siegert shift~\cite{baust2016}, and the emergence of two-mode squeezed vacuum as the ground state~\cite{ciuti2005} in the ultrastrong regime, with a potential for realizing an ultrafast two-qubit control phase gate~\cite{romero2012} in circuit-QED. The experiments in the ultrastrong coupling regime have shown great flexibility in the tunability of parameters, and the ultrastrong coupling far surpasses dissipation-induced decoherence, which typically disrupts the delicate quantum states essential for computation. Thus, the ultrastrong coupling can realize quantum gates with extremely short operation times, enabling the completion of more operations before decoherence compromises the system.
	
	In this paper, we propose an efficient and rapid entangling quantum gate for two or multi-qubit operations within the ultrastrong coupling regime. The protocol leverages the robust, strong, tunable interaction between matter and the bosonic field, allowing for short gate operation times. The gate time is independent of the qubits' initial states, relying solely on the strong coupling dynamics between matter and the bosonic field. The core concept involves linearly coupling two or more matter qubits to a harmonic oscillator formed by the bosonic modes, in a similar way to the M{\o}lmer-S{\o}rensen gate~\cite{sorensen1999,sorensen2000} for trapped ion systems. Consequently, the qubit-field interactions cause the bosonic state to experience a constant force. This force displaces the bosonic field by an amount proportional to the spin of the qubits. The displacement vanishes periodically, thus disentangling with qubit states. Hence, the final state of bosons at these periods differ from its initial state only by a global phase. However, the interactions mediate time-dependent spin-spin interactions in the qubits that persist at the disentangling times. The spin-spin interactions at those times are responsible for entangling the qubit states, thus realizing a quantum gate. We characterize this quantum gate and assess the fidelity of the entangled states it produces.

	The remainder of the paper is organized as follows. Section \ref{sec:CollectiveSpins} describes the operations for manipulating the qubits and the model that is considered in this paper. In Sec. \ref{sec:Rabi}, we present the corresponding unitary operator for entangling spin qubit states. It is followed by an analysis of this operator's properties in Sec. \ref{sec:EntangledStates}. Section \ref{sec:EntangledStatePrep} details the state preparation for even or odd numbers of qubits. The effects of free evolution on the prepared states are explored in Sec. \ref{sec:FreeEvolution}. Section \ref{sec:Experiment} discusses potential experimental implementations, and Sec. \ref{sec:Summary} summarizes our results and conclusions.

		\begin{figure}[t]
				\includegraphics[width = \columnwidth]{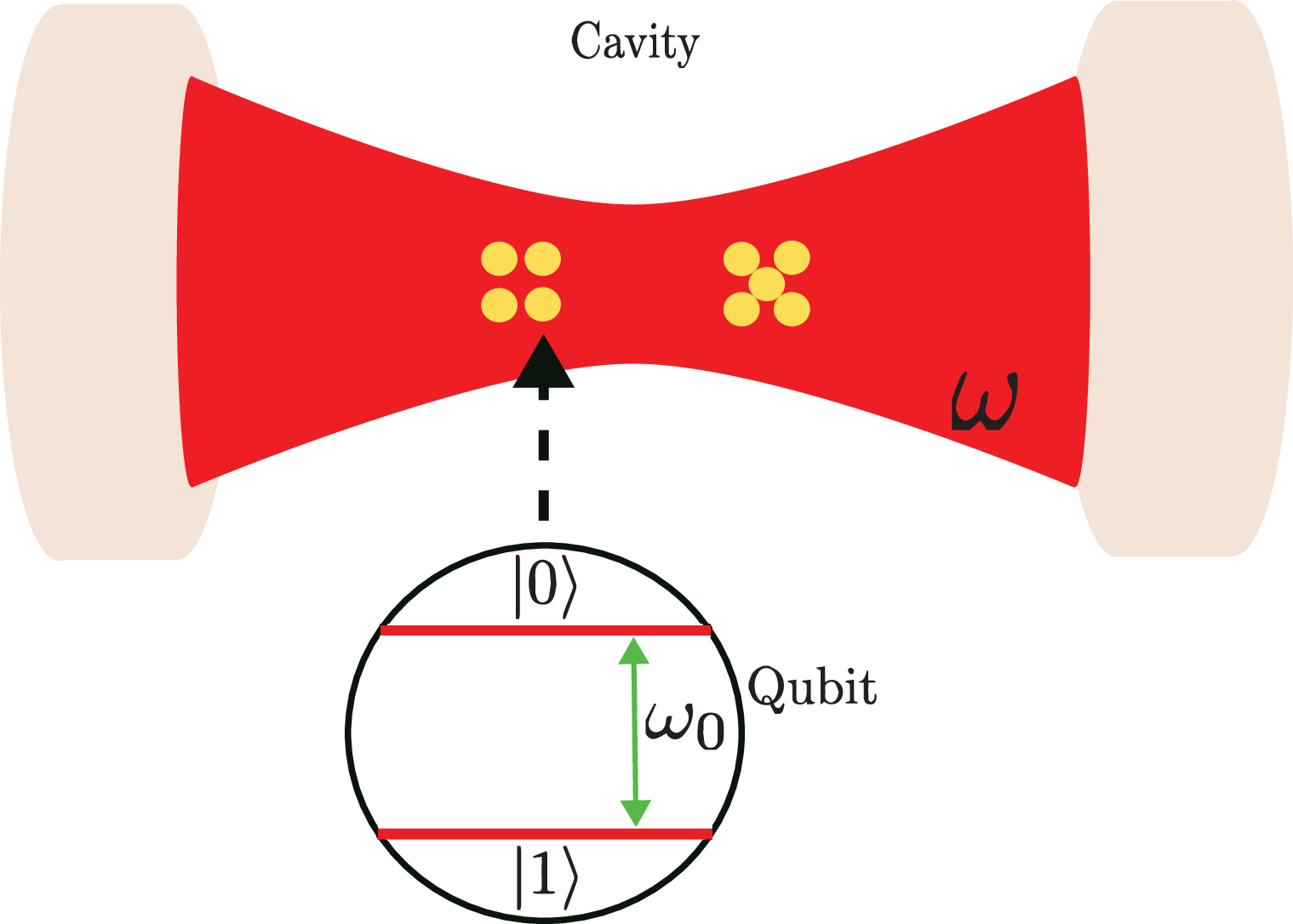}
				\caption{Two qubit ensembles interacting with a common bosonic mode.
    }
				\label{fig1}
			\end{figure}
	
	\section{Physical system and model\label{sec:CollectiveSpins}}
	We first describe the types of collective operators for many-qubit systems that we consider in this paper.  Examples of the system we have in mind are ensembles of neutral atoms in a trap, trapped ions, and superconducting qubits. Consider a system comprised of $ N $ such qubits. Each qubit labeled by the index $ m $ may be described as an effective spin-1/2 system, with operators
 \begin{align}
S^x_m = \frac{1}{2} \left(
\begin{matrix}
0 & 1 \\
1 & 0 \\
\end{matrix}
\right),   
S^y_m = \frac{1}{2} \left(
\begin{matrix}
0 & -i\\
i & 0 \\
\end{matrix}
\right) ,  
S^z_m = \frac{1}{2} \left(
\begin{matrix}
1 & 0 \\
0 & -1\\
\end{matrix}
\right) . 
 \end{align}
We denote the eigenstates of $ J^z_m $ as $ | 0 \rangle, |1 \rangle $, with eigenvalues $ +1/2, -1/2 $, respectively.  The spin operator of the \emph{m}th and \emph{n}th  qubit satisfies the following commutation relation
	\begin{equation}
		\label{eq:CommutationRelation}
		[S^\alpha_m,S^\beta_n] = i\delta_{m,n}\epsilon_{\alpha\beta\gamma}S^\gamma_n,
	\end{equation}
	where $\epsilon_{\alpha\beta\gamma}$ is the Levi-Civita antisymmetric tensor and $\alpha,\beta,\gamma\in\{x,y,z\}$.

 The operators of the $N$ qubit systems can be mapped to a spin-$j$ system where $j = N/2$.  We define the collective spin operators as
	\begin{equation}
		\label{eq:CollectiveOperator}
		J^\alpha = \sum_{n=1}^{N} S^\alpha_n,
	\end{equation}
    where $\alpha \in\{x,y,z\}$. The \emph{z}-projection of the collective spin has eigenstates
	\begin{equation}
		\label{eq:ActionOfZOperator}
		J^z \lvert j, m \rangle = m\lvert j, m\rangle,
	\end{equation}
	where $m$ takes on values $-j,\,-j+1,\ldots,\,j$. 

	The model we consider is as shown in Fig.~\ref{fig1}. We examine \emph{N} qubits interacting with a common bosonic mode. Such qubit systems interacting with bosons model many physical systems, for example, trapped ions~\cite{lv2018}, superconducting qubits~\cite{yoshihara2017}, neutral atoms~\cite{baumann2010,koch2023,hunanyan2024}, a nuclear spin interacting with magnetic field~\cite{rabi1936,rabi1937}, electrons coupled to a phonon mode of a crystal lattice~\cite{holstein1959,agarwal2013}, or an LC circuit~\cite{chiorescu2004,johansson2006}. The dynamics of such a system are described by the Hamiltonian~\cite{rabi1937,barnett2002,agarwal2012} 
	\begin{equation}
		\label{eq:RabiHamiltonian}
		H = \hbar \omega_0J^z + \hbar\omega \hat{a}^\dagger\hat{a}  + \hbar GJ^x(\hat{a}^\dagger + \hat{a}),
	\end{equation}
	where $J^{\alpha}$ is the total spin operator. The parameter $G$ is the qubit-oscillator coupling strength, and $\omega_0$ is the transition frequency of a qubit. The operators $\hat{a}$ and $\hat{a}^\dagger$ annihilate and create quanta, respectively, from the bosonic mode with frequency $\omega$.

	For many applications~\cite{miller2005,walther2006}, the strength of qubit-oscillator coupling $G$ is small compared to the other frequencies in the Hamiltonian~(\ref{eq:RabiHamiltonian}). For these situations, one typically uses the rotating wave approximation~\cite{shore1990v1,walls2008}, where terms such as $ J^+ a^\dagger $ and $ J^- a$ in~(\ref{eq:RabiHamiltonian}) are dropped. However, various experiments~\cite{colombe2007,lv2018,koch2023,hunanyan2024} are steadily moving towards the ultrastrong coupling limit. In this regime, the transition frequency $\omega_0$ is small compared to the cavity mode frequency $\omega$ and the qubit-boson coupling strength $G$. 
 
 Defining a dimensionless time as $\tau = \omega t$, a dimensionless coupling strength $g = G/\omega$, and  a dimensionless qubit frequency $\epsilon = \omega_0/\omega$, we obtain a dimensionless Hamiltonian $H = H/(\hbar\omega)$, and (\ref{eq:RabiHamiltonian}) becomes
	\begin{equation}
		\label{eq:DimensionlessHamitonian}
		H = \epsilon J^z + \hat{a}^\dagger\hat{a} + gJ^x(\hat{a}^\dagger + \hat{a}).
	\end{equation}
		 
	In the following, we assume that the energy splitting $ \epsilon $ of the qubits can be controlled and be set to zero.  For example, in neutral atoms, a Zeeman magnetic field may be applied to suitably chosen hyperfine ground states to put them into degeneracy.  We examine the effect of any residual $ \epsilon $ later in Sec.\ref{sec:FreeEvolution}. Hence, the Hamiltonian that drives the homogeneous solution everywhere is 
	\begin{equation}
		\label{eq:HamiltonianH_0}
		H_0 = \hat{a}^\dagger\hat{a} + gJ^x(\hat{a}^\dagger + \hat{a}) .
	\end{equation}

	\section{Exact solution of ultrastrong coupling model\label{sec:Rabi}}	
    We now show that the ultrasrong coupling model (\ref{eq:HamiltonianH_0}) can be solved exactly. Our approach uses a similar approach to Ref.~\cite{sorensen2000} with the modifications that we write the unitary evolution operator $U=e^{-i H_0\tau}$ as a product of unitary exponential operators in the nonrotating frame of the bosons rather than the rotating frame of the bosonic oscillator, $\hat{a}^\dagger \hat{a}$. 

    The composite state $\lvert\psi\rangle$ of the qubits and bosonic field  evolves according to the Schr\"odinger equation with  Hamiltonian (\ref{eq:HamiltonianH_0}) as
	\begin{equation}
		\label{eq:SchrodingerEquation}
		i\frac{d \lvert \psi\rangle}{d\tau} = H_0 \lvert \psi \rangle. 
	\end{equation}
	The solution of (\ref{eq:SchrodingerEquation}) at any time $\tau$ for a given initial state $\lvert\psi_0\rangle$ is 
	\begin{equation}
		\label{eq:StateSolution}
		\lvert \psi(\tau)\rangle = 
		\hat{U}(\tau) \lvert\psi_0\rangle,
	\end{equation}
	where the unitary operator $\hat{U}(\tau)$ is 
	\begin{equation}
		\label{eq:UnitaryEvolution}
		\hat{U} = e^{-i\tau\left[\hat{a}^\dagger\hat{a} + gJ^x(\hat{a}^\dagger + \hat{a}) \right]}.
	\end{equation}
	The initial state $\lvert\psi_0\rangle$ is a product state of the form 
	\begin{equation}
		\label{eq:InitialStateSystem}
		\lvert \psi_0\rangle = \lvert\psi_{\mathrm{Q}}\rangle\otimes \lvert \psi_{\mathrm{B}}\rangle,
	\end{equation}
	where $\lvert\psi_{\mathrm{Q}}\rangle$ is the initial state of the qubits. The initial state of bosons $\lvert \psi_{\mathrm{B}}\rangle$ is a coherent state
	\begin{equation}
		\label{eq:CoherentState}
		\lvert \psi_{\mathrm{B}}\rangle = e^{-\frac{|\alpha|^2}{2}} \sum _{n=0}^{\infty} \frac{\alpha^n}{\sqrt{n}} \lvert n\rangle,
	\end{equation}	
	where $\alpha$ is the average boson number amplitude. 	

	The action of the unitary operator (\ref{eq:UnitaryEvolution}) on the creation and annihilation operator is as follows (see Appendix~\ref{sec:DisplacementOperator}):
	\begin{align}
		\label{eq:UnitaryOnAnnihilationOperator}
		\hat{U}^{-1}\hat{a}\hat{U} & = \hat{a}e^{-i\tau} - gJ^x\left(1 - e^{-i\tau} \right ),\\
		\label{eq:UnitaryOnCreationOperator}
		\hat{U}^{-1}\hat{a}^\dagger\hat{U} & = \hat{a}^\dagger e^{i\tau} - gJ^x\left( 1 -e^{i\tau} \right ),
	\end{align}
	We see that except for the term $\hat{a}^\dagger\hat{a}$ in (\ref{eq:UnitaryEvolution}) which causes a phase to appear on the operators $\hat{a}$ (on the RHS), the operator $\hat{U}$ displaces the operators $\hat{a}$ and $\hat{a}^\dagger$ much in same way that bears a striking resemblance to the displacement operator. 

	The exponential operator (\ref{eq:UnitaryEvolution}) can be written in an ordered form as a product of exponential operators (see for instance Refs.~\cite{ilo-okeke2023,ilo-okeke2024})
	\begin{equation}
		\label{eq:DisentangledExponentialOperators}
		\begin{split}
			\hat{U}(\tau) = e^{is(\tau)(J^x)^2}e^{ip(\tau)J^x\hat{a}^\dagger}e^{iq(\tau)\hat{a}^\dagger\hat{a}} e^{ir(\tau)J^x\hat{a}},
		\end{split}
	\end{equation}
	where the functions $s$, $p$, $q$, and $r$ are defined as 
	\begin{align}
		\label{eq:SfunctionDisentagledOperator}
		s(\tau) & = g^2\left[\tau + i\left(1 - e^{-i\tau} \right) \right],\\
		\label{eq:PfunctionDisentagledOperator}
		p(\tau) & = ig\left(1 - e^{-i\tau}\right),\\
		\label{eq:QfunctionDisentagledOperator}
		q(\tau) & = -\tau, \\
    	\label{eq:RfunctionDisentagledOperator}
        r(\tau) &= p(\tau).
	\end{align}
	The full derivation is shown in Appendix~\ref{sec:DisentanglingExponentialOperators}. The exponential operators proportional to $\hat{a}J^x$ and $\hat{a}^\dagger J^x$ mediates the entanglement between qubit and boson state. The exponential operator term proportional to $(J^x)^2$ embodies the interaction of the qubits mediated by boson. The interactions thus mediate quantum correlations between qubits.
	
	A special feature of the operator $\hat{U}$ (\ref{eq:DisentangledExponentialOperators}) is that at distinct times $\tau = \tau_d$,
	\begin{equation}
		\label{eq:DisentangledTime}
		\tau_d = 2n\pi,
	\end{equation}
	where $n = 1, \,2,\,3, \cdots$ is a positive integer,  $p(\tau) = r(\tau) =0$, and the quantum states of qubits and bosons disentangle from each other. Since $s(\tau)$ is not zero, the operator $\hat{U}$ at these times consists of nonlinear interactions that generate qubit-qubit correlations, and free evolution of the photon state
	\begin{equation}
		\label{eq:AtomStateAtSpecialPoints}
		\hat{U}(\tau_d) = e^{is(\tau_d) (J^x)^2} e^{iq(\tau_d)\hat{a}^\dagger\hat{a}}.
	\end{equation}  	
	Notice that where $s(\tau_d) = 0$, one recovers the initial state of the qubits.  Consequently, the operator $\hat{U}$ (\ref{eq:DisentangledExponentialOperators}) embodies the nonlinear interactions between qubits mediated by boson. The type of correlations observed depends on the strength of qubit-boson coupling strength. For systems where the qubit-boson coupling strength is weak such as in quantum non-demolition measurements and spin squeezing experiments, the interaction~$(J^x)^2$ is known to give rise to squeezing correlations~\cite{ilo-okeke2021,kuzmich2000}. Where the qubit-boson coupling strength is strong, it produces entangled quantum states such as Schr\"odinger-cat states~\cite{julsgaard2001,aristizabal-zuluga2021,ilo-okeke2021}.

	\section{Two qubit entanglement generation\label{sec:EntangledStates}}
	Here we go beyond the inspection of the unitary operator~(\ref{eq:DisentangledExponentialOperators}) by tracing out the boson states and calculating the purity of the reduced density matrix of the qubits. Because there is entanglement between the qubits and the modes of the boson, the purity is always less than unity except at the times $\tau_d$, when there is no entanglement between the qubits and the bosonic mode. We obtain the relation between the system parameters and the times when purity is maximum.

	\subsection{Purity of qubit states \label{sec:sec:TwoAtomSytem}}
	We first consider two separated spin-$1/2$ qubits interacting with a common bosonic mode. Each qubit is prepared in the state $ \lvert 1 \rangle$ as shown in Fig.~\ref{fig1}. Their combined initial state with the bosons is a product state
	\begin{equation}
		\label{eq:AtomLightInitialState}
		\lvert\psi_0\rangle = \lvert 11 \rangle \otimes\lvert\psi_{\mathrm{B}}\rangle,
	\end{equation}
	where $\lvert \sigma_1 \sigma_2  \rangle = \lvert \sigma_1  \rangle \otimes \lvert \sigma_2 \rangle$ with $ \sigma_1, \sigma_2 \in \{0,1 \} $, $\lvert \psi_{\mathrm{B}}\rangle$ is the initial state of bosons taken as the coherent state (\ref{eq:CoherentState}). Applying the unitary operator (\ref{eq:DisentangledExponentialOperators}) gives the quantum state of the qubit-boson system at any other times as $\lvert \psi(\tau)\rangle = \hat{U} \lvert \psi_0\rangle$. 
	
\begin{figure}[t]
				\includegraphics[width=\columnwidth]{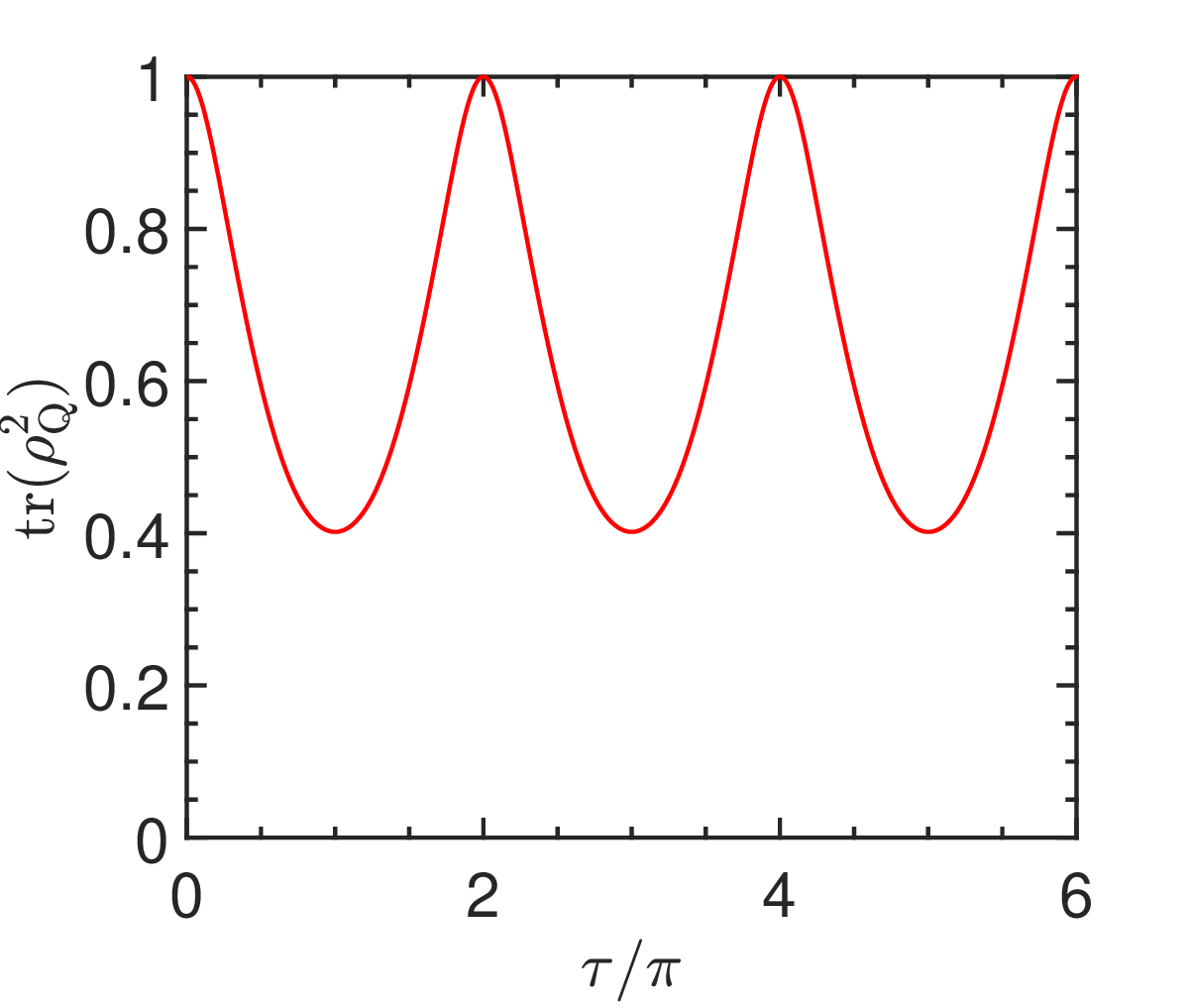}
				\caption{The purity of reduced density matrix of the qubits. For the plot, $g = 1/2$.}
				\label{fig2}
			\end{figure}

	To investigate the disentangling between the quantum states of bosons and qubits, we trace out the quantum state of bosons leaving behind the reduced density matrix of the qubits $\rho_\mathrm{A} = \mathrm{tr}_{\text{B}} (\rho)$, where $\rho = \lvert\psi(\tau)\rangle\langle \psi(\tau)\rvert$ is the density matrix of qubit and bosons, and $\lvert \psi(\tau)\rangle$ is defined in (\ref{eq:StateSolution}). The purity of the qubits quantum state, is then $\mathrm{tr}\left(\rho_\mathrm{A}^2\right)$ and is shown in Fig.~\ref{fig2}. For times $\tau$ not $2\pi$ periodic, the purity of the qubits' reduced density matrix is less than unity showing that both the qubits and bosons are entangled at those times. Thus at these times the unitary operator $\hat{U}(\tau)$ entangles the quantum state of qubits and bosons, and the resulting reduced density matrix of the qubits $\rho_\mathrm{A}$ is generally mixed. However, for times that are $2\pi$ periodic including $0$, the purity is unity and agrees with (\ref{eq:AtomStateAtSpecialPoints}). Hence the quantum states of qubits and bosons are disentangled at those times.

	\subsection{Entanglement Entropy and Entangling Time  \label{sec:sec:sec:EntanglementEntropy}}
	To investigate further the nature of the state obtained at the points where $\tau_d = 2\pi n$, $n = 0,1,2,\,\cdots$, we apply the unitary operator  $\hat{U}(\tau=\tau_d)$ to the initial state $\lvert\psi_0\rangle$. Using (\ref{eq:AtomStateAtSpecialPoints}) gives the state at $\tau= \tau_d$ as
	\begin{equation}
		\label{eq:StateAtDecouplingTime}
		\begin{split}
			\lvert \psi(\tau_d)\rangle = &  \frac{1}{2} \Big[ \left(e^{is} - 1\right)\lvert 00\rangle + \left(1 + e^{is}\right) \lvert 11 \rangle \Big]  \\
			& \otimes e^{iq\hat{a}^\dagger\hat{a}}\lvert\psi_{\mathrm{B}}\rangle .
		\end{split}
	\end{equation}
Note that $e^{iq\hat{a}^\dagger\hat{a}}\lvert\psi_{\mathrm{B}}\rangle$ is the photon state that is different from the initial state by a phase evolution $e^{q\hat{a}^\dagger\hat{a}}$ at time $\tau_d$. Since at times $t= \tau_d$ the state is a product state, tracing out the bosonic states does not affect the purity of the qubit states.

	We readily see from (\ref{eq:StateAtDecouplingTime}) that the function $s$ plays a role in the type of the final state that emerges at $\tau =\tau_d$. For instance if $s=0,$ or $2\pi$, one obtains the initial state of the qubit $\lvert 11 \rangle$. Similarly, $s = \pi$  gives the qubit state  $\lvert 00 \rangle$. Any other value of $s$ between these two values possesses entanglement. Hence it becomes obvious that it is possible to engineer the type of entanglement that appears between the qubits at time $\tau_d$. 
	
	
	The entanglement of a bipartite pure state may be quantified using the von Neumann entropy $S_1 = -\mathrm{tr}(  \rho_1\log_2\rho_1 )$ \cite{nielsen2000,byrnes2021}, where $ \rho_1 = \text{tr}_{B,2} (|\psi(\tau_d) \rangle \langle \psi(\tau_d)  | ) $ is the reduced density matrix of qubit 1.  This evaluates as
	\begin{align}
		\label{eq:EntanglementEntropy}
		S_1 = &-\frac{(1 - \cos s)}{2}\log_2\left(\frac{1 - \cos s}{2}\right) \nonumber\\
		& -\frac{(1 + \cos s)}{2}\log_2\left(\frac{1 + \cos s}{2}\right).
	\end{align}
	We immediately see from the entanglement entropy (\ref{eq:EntanglementEntropy}) that there exists some entanglement or correlation between qubits 1 and 2 for $s \neq m\pi$, where $m = 0,\, 1,\,2,\,3,\,\ldots$. At specific $s$ values and odd multiples of $\pm \pi/2$, the entanglement entropy reaches its maximum value. Thus, at $s = \pm  m_{\mathrm{odd}}\pi/2$ where $m_\mathrm{odd} = 1,\,3,\,5,\,\ldots$ the state prepared at $\tau = \tau_d$ becomes maximally entangled. 

In many quantum information applications, the aim is to produce maximal entanglement (e.g. a CNOT gate).  The above conditions for maximal entanglement can be used to give a condition for the dimensionless qubit-boson coupling $g $.  
Considering positive $s$, $s=m_\mathrm{odd}\pi/2$, and substituting $\tau_d = 2\pi n$ (\ref{eq:DisentangledTime}) in $s$ (\ref{eq:SfunctionDisentagledOperator}), gives $s = g^2\tau_d = 2g^2\pi n$. Solving $m_\mathrm{odd}\pi/2= 2g^2\pi n$  gives $g^2 = m_\mathrm{odd}/(4n)$. Thus, by balancing the ratio $m_\mathrm{odd}/n$, one can tune the interaction strength $g$ for any real number $g \in \mathbb{R}$. Choosing $\tau$ to determine the higher values of $s$, we set $m_\mathrm{odd} = 1$. For $n=1$, the coupling strength becomes $g = 1/2$ so that the maximal entangled state would then appear for times 
	\begin{equation}
		\label{eq:EngineeredEntanglementTime}
		\tau_{\mathrm{Ent}} = 2\pi n_{\mathrm{odd}}.
	\end{equation}
	At other times $\tau_d = n_{\mathrm{even}}\,2\pi$, $n_\mathrm{even}$ takes on even numbers only, the state of the qubits and bosons decouple, and the quantum state of the qubits are not entangled. Choosing a different value of $n$, say $n=2$, gives $g = 1/\sqrt{8}$, with a maximal entangled qubit state appearing at the times $\tau_{\mathrm{Ent}} = 4\pi n_\mathrm{odd}$, where $n_\mathrm{odd} = 1,\,3,\,5,\,7,\ldots $. However, this later entangling time is slower than that given in  (\ref{eq:EngineeredEntanglementTime}), and such slow entangling times will not be discussed further. 
	
	Note that the choice of $s= \pi/2$ always results in a maximal entangled state. Choosing $s$ to be some other value less than $\pi/2$  still prepares a correlated state at the time $\tau_d$ (\ref{eq:DisentangledTime}). Since $s$ determines the type of correlation at $\tau_d$, we can set $s = \pi\beta$, where $\beta $ is some positive real number in the interval $[0,1/2]$, $0\leq\beta\leq 1/2$. Thus, the coupling strength $g$ at time $\tau_d$ becomes
	\begin{equation}
		\label{eq:InteractionStrength}
		g = \sqrt{\frac{\beta}{2\, n}}. 
	\end{equation}

	\subsection{Two Qubit Spin State Entanglement\label{sec:sec:TwoSpins}}

    We now explicitly write the states of the qubits at the maximally entangled times.  
	To prepare a maximally entangled state $g = 1/2$ which gives $s = \pi/2$ and the state of qubits is
	\begin{equation}
		\label{eq:StateAtDecouplingTimeTraced}
		\lvert \psi(\tau =2\pi)\rangle = \frac{e^{i\frac{\pi}{4}}}{\sqrt{2}}\left(i\lvert 00 \rangle + \lvert 11 \rangle \right).
	\end{equation}
	Hence the state at $\tau = 2\pi$ in Fig.~\ref{fig2} corresponds to the state (\ref{eq:StateAtDecouplingTimeTraced}). Similarly, at $\tau= 6\pi$, $s = 3\pi/2$, $p=0$, and (\ref{eq:StateAtDecouplingTime}) reduces to 
	\begin{equation}
		\label{eq:StateAt6pi}
			\lvert \psi(\tau=6\pi)\rangle = \frac{e^{-i\frac{\pi}{4}}}{\sqrt{2}}\left(-i\lvert  00 \rangle + \lvert 11 \rangle \right).
	\end{equation}
	Thus, the state at $\tau = 6\pi$ in Fig.~\ref{fig2} is (\ref{eq:StateAt6pi}). At any other time given by (\ref{eq:EngineeredEntanglementTime}), the entangled state would be either (\ref{eq:StateAtDecouplingTimeTraced})  or (\ref{eq:StateAt6pi}). The states (\ref{eq:StateAtDecouplingTimeTraced}) and (\ref{eq:StateAt6pi}) are maximally entangled as given by the von Neumann entropy.

	\section{Single Ensemble Squeezing and Non-Gaussian states\label{sec:EntangledStatePrep}}
	The interactions $(J^x)^2$ of (\ref{eq:DisentangledExponentialOperators}) are known to generate entanglement between the qubits within the ensemble depending on the qubit-field interaction strength $g$. Applied to a single ensemble, in the short-time regime the interactions create squeezed states~\cite{kitagawa1993,wineland1992,sorensen2001,gross2010,riedel2010}. Here, we engineer quantum entanglement by tuning the qubit-boson interaction strength $g$. We calculate the fidelity of the prepared state for a maximally entangled state. In the case of a non-Gaussian correlated state, such as squeezed states, we calculate the \emph{Q} function of the state and visualize it on the Bloch sphere. 
	
	We have shown that controlling $g$ determines the type of correlations that emerge at $\tau = 2\pi$, since $p =0$ and  (\ref{eq:DisentangledExponentialOperators}) reduces to (\ref{eq:AtomStateAtSpecialPoints}). Thus the quantum state of qubit and bosons is $\lvert \psi (\tau_d) \rangle = \hat{U}(\tau_d)\lvert\psi_0\rangle$. Tracing out the quantum states of bosons gives the state of the qubits as 
	\begin{equation}
		\label{eq:AtomStateAtDecouplingTime}
		\begin{split}
			\lvert \psi(\tau_d)\rangle & = e^{s(J^x)^2}\lvert \psi_{\mathrm{Q}}\rangle.
		\end{split}
	\end{equation}
	In the following, we first analyze this state for maximal state preparation using different numbers of particles classified as even or odd, beginning with two particles. Later, we also analyze the state~(\ref{eq:AtomStateAtDecouplingTime}) for non-maximal fidelity quantum state preparation.

	\subsection{Many-particle Entangled State\label{sec:sec:ManyParticleState}}
	In Sec.~\ref{sec:sec:TwoSpins}, we showed that the $J^2_x$ interaction produces the maximally entangled state $\lvert\psi(\tau = 2\pi)\rangle$. The quantum state $\lvert\psi(\tau = 2\pi)\rangle$ is the lowest order GHZ state where all the particles are in a linear superposition of state $\lvert0 \rangle^{\otimes N} $ and $\lvert 1 \rangle^{\otimes N}$.  Here, we extend this idea to prepare an \emph{N}-particle GHZ-like state ($N\ge 3$) with unit fidelity from \emph{N} particles initially prepared in their ground state $\lvert 1\rangle^{\otimes N}$.

	\begin{figure}[t]
		\includegraphics[width=\columnwidth]{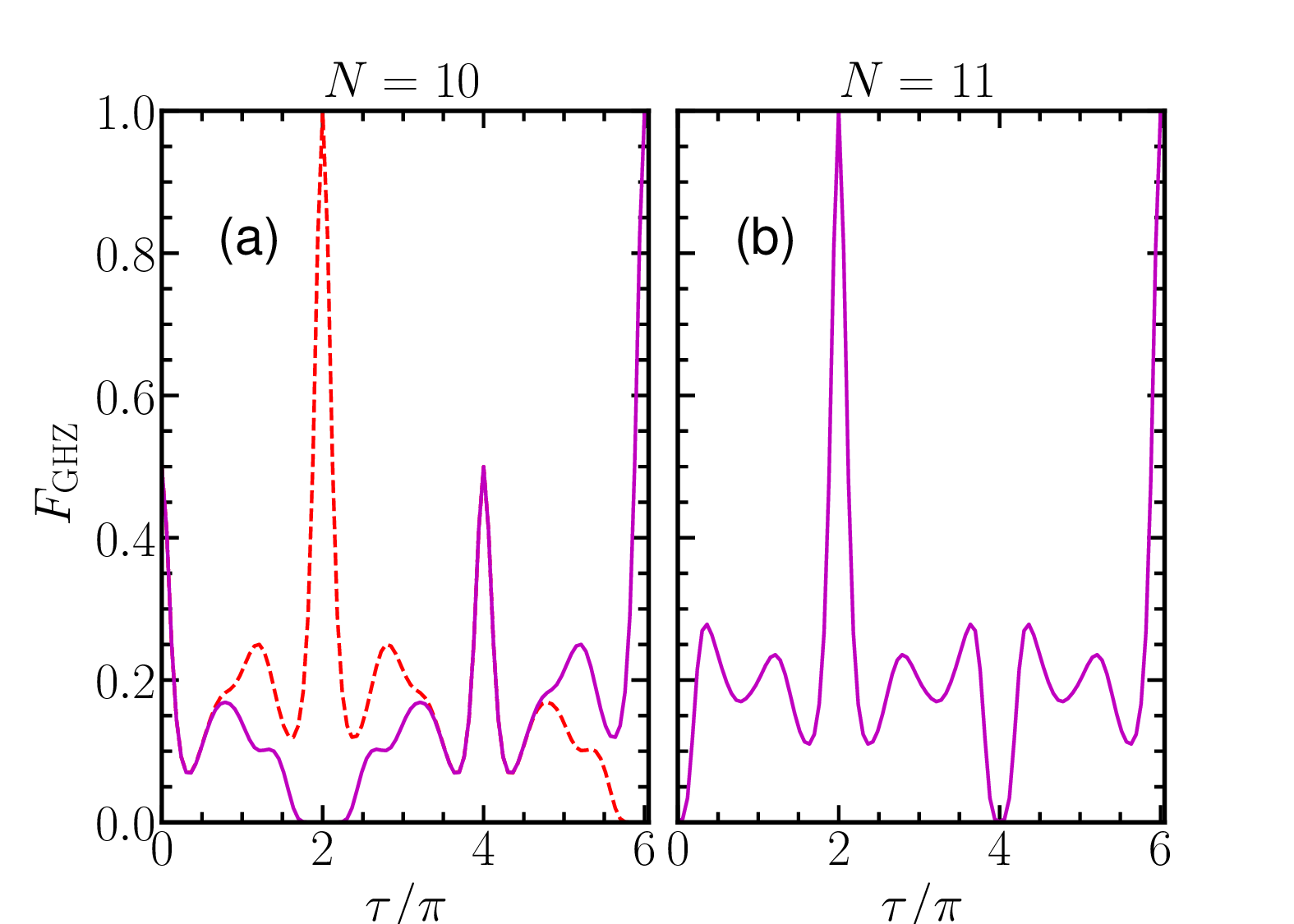}
		\caption{Evolution of the Fidelity (\ref{eq:GHZFidelity}) of the GHZ state for $N = 10$ and $11$ spin-$1/2$ particles. The qubit-boson interaction $g = 1/2$. The initial state of bosons is a coherent state~(\ref{eq:CoherentState}) with the average boson number amplitude $\alpha = \sqrt{2}$. 
		}
		\label{fig3}
	\end{figure}

	\begin{figure*}[!]
		\includegraphics[width=\textwidth,keepaspectratio]{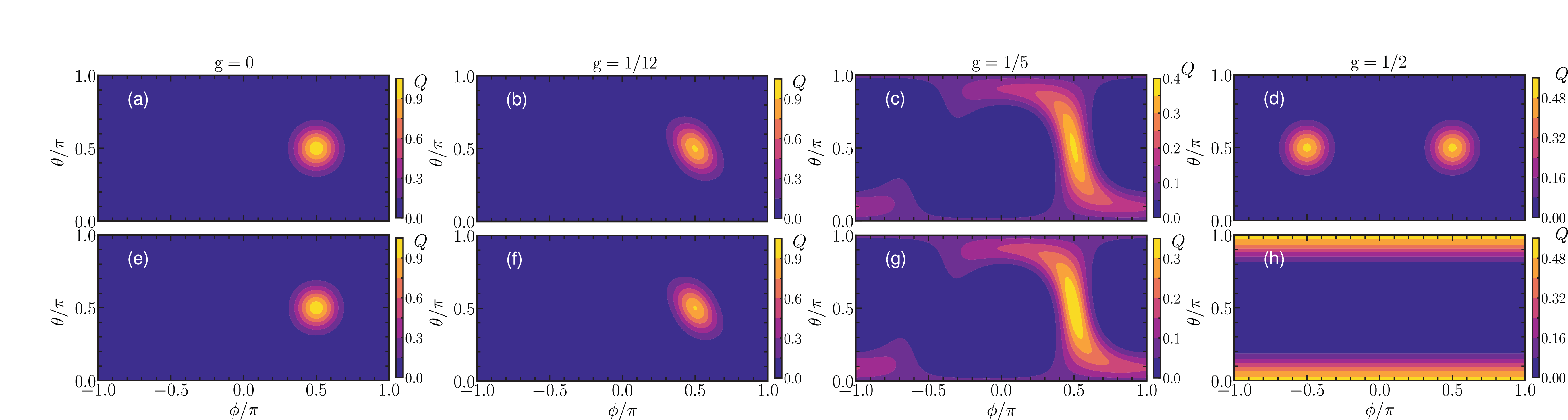}
		\caption{The \emph{Q} function of the quantum state of the qubits at $\tau_{\mathrm{Ent}} = 2\pi$ for varying strengths of the qubit-boson interaction strength $g$. The first row contains $N=20$ spins-$1/2$ particles, while the second row contains $N=21$ spin-$1/2$ particles.}
		\label{fig4}
	\end{figure*}
	
	\subsubsection{Even Number Of Qubits \label{sec:sec:sec:EvenNumberParticle}}
	We first consider the case where the number of qubits \emph{N} is even. With the interaction strength $g$ set to $1/2$, and up to some global phase, either one of the GHZ states would always appear at times given by (\ref{eq:EngineeredEntanglementTime}),
    \begin{equation}
        \label{eq:GHZStateEven}
        \lvert \psi^{N_\mathrm{even}} (\tau = \tau_\mathrm{Ent})\rangle = \frac{(-1)^{\frac{N + n_\mathrm{odd} + 1}{2}}i \lvert 0\rangle^{\otimes N} + \lvert 1 \rangle^{\otimes N}}{\sqrt{2}},
    \end{equation}
    where $\tau_\mathrm{Ent}$ and $n_\mathrm{odd}$ are defined in (\ref{eq:EngineeredEntanglementTime}).

	To investigate the emergence of a GHZ state in the system, we plot in Fig.~\ref{fig3} the evolution of the fidelity \emph{F}\textsubscript{GHZ} defined as 
	\begin{equation}
		\label{eq:GHZFidelity}
		F_\mathrm{GHZ} = \langle \psi^{N_k}  \lvert \rho_{\text{A}} \lvert \psi^{N_k} \rangle   ,
	\end{equation}	  
	where $k$ is even or odd, $ \rho_{\text{A}} = \text{tr}_B \lvert\psi(\tau)\rangle \langle \psi(\tau) \lvert $ and $\lvert\psi(\tau)\rangle$ is defined in (\ref{eq:StateSolution}).
	
	For an even number of qubits, as shown in Fig.~\ref{fig3}(a), there exist two forms of the GHZ state as given in (\ref{eq:GHZStateEven}). The one observed depends on \emph{N} and \emph{n}\textsubscript{odd} at the time $\tau_{\mathrm{Ent}}$ (\ref{eq:EngineeredEntanglementTime}). For instance, in Fig.~\ref{fig3}(a), since $(N+n_\mathrm{odd}+1)/2$ is even for $n_\mathrm{odd}=1$, the GHZ state that appears at $\tau = 2\pi$ is of the form $\lvert\psi^{N_\mathrm{even}}\rangle\propto i\lvert 0\rangle^{\otimes N} + \lvert 1\rangle^{\otimes N}$. Similarly, for $n_\mathrm{odd}=3$, the GHZ state that emerges at $\tau = 6\pi$ is $\lvert\psi^{N_\mathrm{even}}\rangle\propto -i\lvert 0\rangle^{\otimes N} + \lvert 1\rangle^{\otimes N} $. For a given \emph{N}  and higher $n$ values, the  GHZ state will alternate between these forms as time progresses. The existence of two forms of the GHZ state for $N_\mathrm{even}$ is quite similar to the case for two qubits discussed in the previous section. Note that for $N=2$ in (\ref{eq:GHZStateEven}) we recover (\ref{eq:StateAtDecouplingTimeTraced}) and (\ref{eq:StateAt6pi}), respectively.

	\subsubsection{Odd Number Of Qubits \label{sec:sec:sec:OddNumberParticle}}
	For an odd number of qubits, $\hat{U}(2n_{\mathrm{odd}}\pi)\lvert 1 \rangle^{\otimes N_{\mathrm{odd}}}$, is the superposition of two macroscopic states at $+y$ and $-y$ axis on the Bloch sphere. To observe a maximal fidelity state, in the \emph{z} basis, the state is rotated about the \emph{x} axis by applying the unitary $e^{i\frac{\pi}{2} J^x}$, $\lvert \psi^{N_\mathrm{odd}}\rangle = e^{i\frac{\pi}{2} J^x} \hat{U}(\tau = \tau_{\mathrm{Ent}})\lvert 1 \rangle^{\otimes  N_{\mathrm{odd}}} $. In general, after a rotation of the state $\hat{U}(\tau_{\mathrm{Ent}})\lvert 1 \rangle^{\otimes N}$ about the \emph{x}-axis, the state that emerges at any time is of the form 
	\begin{equation}
		\label{eq:OddNStateAfterRecomb}
		\lvert \psi^{N_\mathrm{odd}}\rangle =e^{i\frac{\pi}{2}J^x} \frac{(-1)^{\frac{(N-3)}{2}}i\lvert 0\rangle^{\otimes N } + \lvert 1 \rangle^{\otimes N}}{\sqrt{2}}.
	\end{equation}
	The evolution of the GHZ state is investigated using the fidelity  \emph{F}\textsubscript{GHZ} (\ref{eq:GHZFidelity}) where $k$ is odd.

	In contrast to an even number of particles, only one GHZ-like state exists for an odd number of qubits since there is no dependence on the \emph{n}\textsubscript{odd}. Fig.~\ref{fig3}(b) shows the evolution of the fidelity (\ref{eq:GHZFidelity}) for an odd number of qubits. At times other than $\tau_{\mathrm{Ent}}$~(\ref{eq:EngineeredEntanglementTime}), there is dissimilarity between the state at time $\tau$, $\lvert\psi(\tau)\rangle$ (\ref{eq:StateSolution}) and (\ref{eq:OddNStateAfterRecomb}).  Hence, the fidelity (\ref{eq:GHZFidelity}) is not unity at those times. However, at times $\tau=\tau_{\mathrm{Ent}}$ (i.e. $\tau = 2\pi$ and $\tau=6\pi$) the state $\lvert\psi(\tau)\rangle$ and  (\ref{eq:OddNStateAfterRecomb}) are the same thus resulting in unit fidelity as seen in Fig.~\ref{fig3}(b).

 	\subsection{Squeezed and Non-Gaussian States \label{sec:SqueezedStatePrep}}
 	In this section, we show other types of states, such as squeezed states, can appear at $\tau_{\mathrm{Ent}}$ using large \emph{N} particles. The quantum state prepared at $\tau = \tau_{\mathrm{Ent}}$ is (\ref{eq:AtomStateAtDecouplingTime}), $\lvert\psi\rangle = e^{s(J^x)^2}\lvert \psi_{\mathrm{Q}}\rangle$ where $\lvert\psi_{\mathrm{Q}}\rangle$ is the initial state of the qubits. 
 	
 	To see the effect of the induced total spin interactions among the spins, we calculate the \emph{Q} function of the prepared state. The initial state of the spins points along the positive \emph{y} axis on the Bloch sphere. This state has its noise distributed symmetrically about the mean spin direction and is characteristic of a Gaussian distribution (Figs. ~\ref{fig4}(a)(e)).  The initial state is no longer circular for values of $g$ different from zero. It is squeezed in a diagonal direction (Figs. ~\ref{fig4}(b)(f)). 
    The squeezing heralds the presence of entanglement between the particles \cite{sorensen2001many}, which are not present at $g=0$. The amount of the correlations increases with $g$, resulting in non-Gaussian states. This is visible by the deformation of distribution giving a non-Gaussian shape about its mean direction, as shown in Figs.~\ref{fig4}(c) and~(g)~\cite{ilo-okeke2021,ilo-okeke2023}.
  
  At $g = 1/2$, a Schr\"odinger cat state emerges as before (Sec.~\ref{sec:EntangledStatePrep}). However, for an even number of spins, the Schr\"odinger cat state emerges as a superposition of macroscopic state along the $\pm y$ axis. On the other hand, for an odd number of spins, the Schr\"odinger cat state emerges as a superposition of macroscopic state along the $\pm z$ axis. These are the reverse of the results obtained in Sec.~\ref{sec:EntangledStatePrep} where the initial state of the qubits were in the negative \emph{z} direction.  Hence varying qubit-boson interaction strength $g$, or entangling time $\tau_{\mathrm{Ent}}$, or $s$ shows that one can prepare different quantum states. This bears a resemblance to generating different quantum states of macroscopic spin state at different gate times~\cite{byrnes2013,gao2022}.

    \section{Multiple Ensemble Entanglement \label{sec:MultipleEnsembles}}
    Here we investigate the generation of squeezing and entanglement in more an one ensemble. With multiple ensembles, we expect that entanglement would be generated between the ensembles ~\cite{julsgaard2001,fadel2018,kunkel2018,lange2018,ilo-okeke2021,aristizabal-zuluga2021}. A full derivation of the many-qubit ensembles interacting with a common light mode is discussed in Appendix~\ref{sec:EnsembleCase}.  But we may understand the nature of the entanglement that is generated by noting that we may write the total spin of multiple ensembles as
\begin{align}
J^\alpha =  \sum_{k=1}^M J^\alpha_k
\end{align}
where $ M $ is the number of ensembles.  Each of the ensembles are labeled by the index $ k $, and consist of $ N_k $ qubits, with a total spin $ N_k/2$, which can be described by the operator $ J^\alpha_k $. The collective spin operators of the \emph{k}\textsuperscript{th} ensemble are defined in a similar way to (\ref{eq:CollectiveOperator}).  Viewing the multi-ensemble system as a single ensemble, we may carry over our previous results and we obtain an evolution (\ref{eq:AtomStateAtDecouplingTime}).  The square of the $ J^x $ operator contains cross-terms between the ensembles, which produces a one-axis two-spin (1A2S) squeezing \cite{byrnes2013}.  For example, for $ M = 2$ ensembles the interaction takes the form
\begin{align}
(J^x)^2 & = (J^x_1 + J^x_2)^2 \nonumber \\
& = (J^x_1)^2 + (J^x_2)^2 + 2 J^x_1 J^x_2 .
\end{align}
The first two terms describe individual squeezing on the two ensembles, and the last term is an entangling term of the 1A2S form.  

We now describe in more detail the $ M = 2 $ ensemble case. Each \emph{k}\textsuperscript{th} ensemble interacts only with a common bosonic mode with an interaction strength $g_1,\, \, g_2$. Suppose that the initial state of each ensemble is $\lvert \psi_{\text{Q}} \rangle_k$ while the bosons are in the state $\lvert\psi_{\text{B}}\rangle$. The unitary operator $\hat{U}$ that takes any initial state $\lvert\psi_0\rangle = \lvert\psi_{\text{Q}}\rangle_1 \otimes \lvert\psi_{\text{Q}} \rangle_2 \otimes\lvert\psi_{\text{B}} \rangle$ of the qubit ensembles and bosons to a state $\lvert\psi(\tau)\rangle = \hat{U}(\tau)\lvert\psi_0\rangle$ at time $\tau$ (\ref{eq:StateSolution}) is 
    \begin{equation}
        \label{eq:EnsembleUnitary}
         \hat{U}(\tau) = e^{-i\tau \left[ \hat{a}^\dagger \hat{a} + g J^{x}(\hat{a}^\dagger + \hat{a})\right]},
    \end{equation}    
    where $g =\sqrt{g^2_1+ g_2^2}$ is the magnitude of qubit-boson interaction strength, the effective spin $J^x = J^x_1 \cos\theta  + J^x_2\sin\theta $ is the sum of the ensembles \emph{x}-spin operators, and $\cos\theta = g_1/g$ and $\sin\theta = g_2/g$. The parameter $\theta$ defines the ratio $g_2/g_1$ of the ensemble-boson interaction strength and lies in the interval $-\pi/4\leq \theta\leq \pi/4$. It determines the mixing of the ensemble spins via $(J^x)^2$ term of~(\ref{eq:DisentangledExponentialOperators}). Here, the first ensemble is the ensemble whose magnitude of the qubit-boson interaction strength is greater than the other and denotes its interaction as $g_1$. 
    
    One ensemble is present for the values $\theta = 0$, and we recover the one ensemble case discussed in Sec.~\ref{sec:EntangledStatePrep}. However, for other values of $\theta$, the qubit-boson coupling mediates interactions between the qubit ensembles and entangles them. Note that the unitary (\ref{eq:EnsembleUnitary}) is of the same form as (\ref{eq:UnitaryEvolution}). As such, the state of the ensembles and bosons decouple at times (\ref{eq:DisentangledTime}). Similarly, the purity of the ensemble states obtained after tracing out the bosonic state is the same as that shown in Fig.~\ref{fig2}. Quantifying the entanglement between the ensembles would depend on the ratio of the ensemble-boson interaction strength $g_2/g_1$ and the magnitude of their couplings $g$. 

    To investigate the inter-ensemble entanglement generated between the ensembles, we use the product state $\rho_0 = \lvert \psi_0\rangle\langle\psi_0\rvert  $ as an initial state to the unitary gate (\ref{eq:EnsembleUnitary}), then calculate the negativity 
    \begin{equation}
        \label{eq:Negativity}
        \mathcal{N}(\rho_{\text{Q}}) = \frac{\lvert\lvert \rho_{\text{Q}} ^{\mathrm{T}_2}\rvert \rvert -1}{2},
    \end{equation}
    where $\rho_{\text{Q}} = \mathrm{tr}_{\text{B}}(\lvert\psi(\tau)\rangle\langle\psi(\tau)\rvert)$ is the reduced density matrix of the qubit ensembles, $\rho_{\text{Q}} ^{\mathrm{T}_2}$ is the partial transpose of the reduced density matrix $\rho_{\text{Q}}$ of the qubit ensembles with respect to the ensemble $2$, and $\lvert\lvert\,\cdot\,\rvert\rvert$ is the trace norm. 
    
\begin{figure}[t]
		\includegraphics[width=\columnwidth]{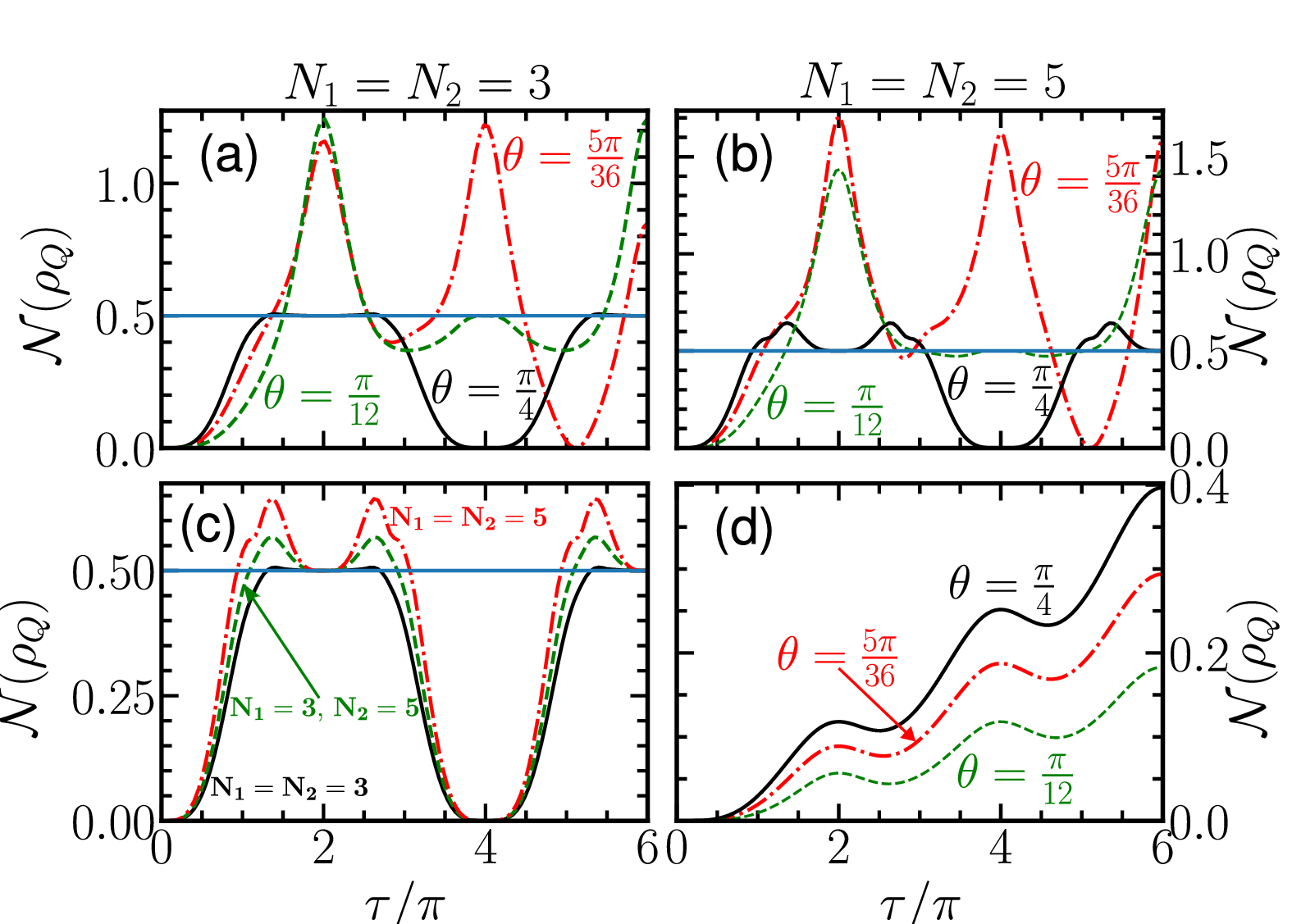}
		\caption{The evolution of the negativity of two ensemble qubits. In (a) - (c), the magnitude of the coupling parameter $g = 1/\sqrt{2}$ while in (d), the magnitude of the coupling parameter is $g = 1/(6\sqrt{2})$. (b) and (d) have the same ensemble qubit populations as indicated above (b). (c) is the negativity at different ensemble qubit populations for $\theta = \pi/4$. The initial state of bosons is a coherent state (\ref{eq:CoherentState}) with the average boson number amplitude $\alpha = \sqrt{2}$. The solid line at negativity $\mathcal{N}(\rho_A) = 0.5$ is a guide for the eye.
			}
		\label{fig5}
		\end{figure}

    We limit our investigation to the interval $0\leq\theta\leq\pi/4$ since the interactions are symmetric. The evolution of negativity $\mathcal{N}(\rho_{\text{Q}})$ is shown in Fig.~\ref{fig5}. A single ensemble, which corresponds to $\theta = 0$, has zero negativity. However, as the $\theta$ value increases, the negativity increases with time, thus signaling the presence of entanglement in the qubit-boson system. For weak qubit-boson coupling strengths, $\lvert g_2\rvert/\lvert g_1\rvert \ll 1 $ and $g \ll 1$ shown in Fig.~\ref{fig5}(d), the negativity increases with time. However, there is a hump at the times where the qubit ensemble and bosons are disentangled $\tau = \tau_d$ (\ref{eq:DisentangledTime}).  

    In Figs.~\ref{fig5}(a)-(c), we chose $g$ such that for the ratio of the interactions $\lvert g_2\rvert/\lvert g_1\rvert$ equal to unity, the effective interaction strength is one-half, $gJ^x = (J^x_1 + J^x_2)/2$. At the special times (\ref{eq:DisentangledTime}) when the qubit ensemble and the bosons disentangle, the negativity reaches a peak value. At these times, entanglement initially residing between the qubits and bosons now concentrates entirely within the qubit ensembles. As the qubit-boson couplings become equal in magnitude $\theta = \pm\pi/4$, there is a suppression of negativity peaks at some disentangling times (\ref{eq:DisentangledTime}) due to the destructive interference resulting from interactions between the qubit ensembles. For instance, with $N_1= N_2 = 3$ in Fig.~\ref{fig5}(a), the peak at $\tau = 4\pi$ for $\theta = 5\pi/36$ is completely destroyed at $\theta = \pi/4$, while the negativity peaks at $\tau=2\pi$ and $\tau=6\pi$ are leveled to a plateau, respectively. A similar effect is seen for $\theta = 5\pi/36$ where the peak at $\tau = 4\pi$ is flattened to a plateau at  $\theta = \pi/12$. Similar observations are made in Fig.~\ref{fig5}(b) for $N_1 = N_2 = 5$. Note that these plateaus tend to a particular value $\mathcal{N}(\rho_A) = 1/2$.

    \subsection{Equal coupling strengths }
    To better understand the behavior of negativity, we take the couplings to be the same $g_1= g_2 $ so that $\theta = \pi/4$ and we look at the variation of the negativity with the number of qubits in each ensemble, $N_1$ and $N_2$. Then, $J^x = J^x_1 + J^x_2$ of (\ref{eq:EnsembleUnitary}) has the same form as the one ensemble case. As such, the effective qubit-boson interaction strength is one-half of the total for each ensemble. The ensembles are in two different physical locations. Fig.~\ref{fig5}(c) shows the negativity of this case. The increasing number of qubits in either ensemble causes the plateaus to shrink to a point at $\tau= 2\pi$ and $6\pi$. This becomes a narrow or sharp fault line between the two peaks in the limit of a large number $N$ of qubits. As we already saw in Sec.~\ref{sec:EnsembleCase}, the state at this time $\tau = 2\pi$ or $6\pi$ is a Schr\"odinger cat state. Hence, one prepares a Schr\"odinger cat state between the qubit ensembles when $\tau_d = \tau_\mathrm{Ent}$ (\ref{eq:EngineeredEntanglementTime}). For other times $\tau_d$ not equal to the entangling time $\tau_d \neq \tau_\mathrm{Ent}$, the negativity is zero since there is no entanglement at those times, $\mathcal{N}(\rho_A)=0$ as shown at $\tau = 4\pi$ in Fig.~\ref{fig5}(c). 

    \subsection{Unequal coupling strengths }
    This behavior sheds light on the available entanglement where the ensemble qubit-boson interaction strengths are not identical. For example, in Figs.~\ref{fig5}(a) and (b) for $\theta= 5\pi/36$, varying the number of qubits in each ensemble causes a dip or minimum in negativity between the peaks at $\tau = 2\pi$ and $\tau = 4\pi$. The minimum between these two peaks tend toward $\mathcal{N}(\rho_A) = 1/2$ at a time when the quantum states of the qubit ensembles and boson field are entangled. The qubit ensemble equally disentangles with the bosonic field. However, the disentangling time $\tau_d$ for  $\theta = 5\pi/36$ occurs at the rational time $\tau_d \approx 5.108\pi$ that is not easily predicted by (\ref{eq:DisentangledTime}). The same analysis applies to the negativity calculated at qubit-boson coupling ratio $\theta =\pi/12$ shown in Figs.~\ref{fig5}(a) and (b). In this instance, the disentangling time occurs beyond the time window shown. Also, note that the cat state is emerging in the neighborhood of $\tau=4\pi$ for $\theta = \pi/12$ which is not there in the single ensemble case.  
    
    More generally, for significant identical coupling strengths, the negativity for a large number of qubits tends to a fixed value $\mathcal{N}(\rho_A) = 1/2$ at $\tau_d = \tau_\mathrm{Ent}$ (\ref{eq:EngineeredEntanglementTime}) where the state at those times are known to be a Schr\"odinger cat state. At other times, $\tau_d \neq \tau_\mathrm{Ent}$, there is no entanglement between the qubit ensembles, $\mathcal{N}(\rho_A) = 0$. And where the interaction strength is not identical, the cat state could appear at the times where the quantum states of qubit ensemble and light are entangled while the disentangling time could become a rational number. Note that for a given number of qubits in each ensemble with dissimilar interaction strengths $\lvert g_2\rvert \neq\lvert g_1\rvert$, the available entanglement at the time $\tau_\mathrm{Ent}$ (\ref{eq:EngineeredEntanglementTime}) is more than that with the same interaction strength $\lvert g_2\rvert = \lvert g_1\rvert$. The enhancement comes about due to constructive interference from the ensembles' interactions at the time $\tau_\mathrm{Ent} $ compared to the destructive interference when ensembles have the same interaction strength, $\lvert g_2/g_1\rvert = 1$. Thus, at the time $\tau_\mathrm{Ent} $, the negativity of the ensembles with dissimilar interaction strengths is more than those with the same interaction strength.

	\section{Effect Of Free Qubit Evolution\label{sec:FreeEvolution}}
	In this section, we account for the free oscillation of the qubits  $\epsilon J^z$ by moving into the interaction picture of $H_0$. We consider a situation that the qubits can be made nearly degenerate, such that $ \epsilon \ll 1 $.  It is however important to understand what the effect of having a residual $ \epsilon $ is.  Using the purity and fidelity of the maximally entangled states, we study the impact of free oscillation term on the state preparation at the weak and significantly strong free oscillation parameters.

\begin{figure}[t]
		\includegraphics[width=\columnwidth]{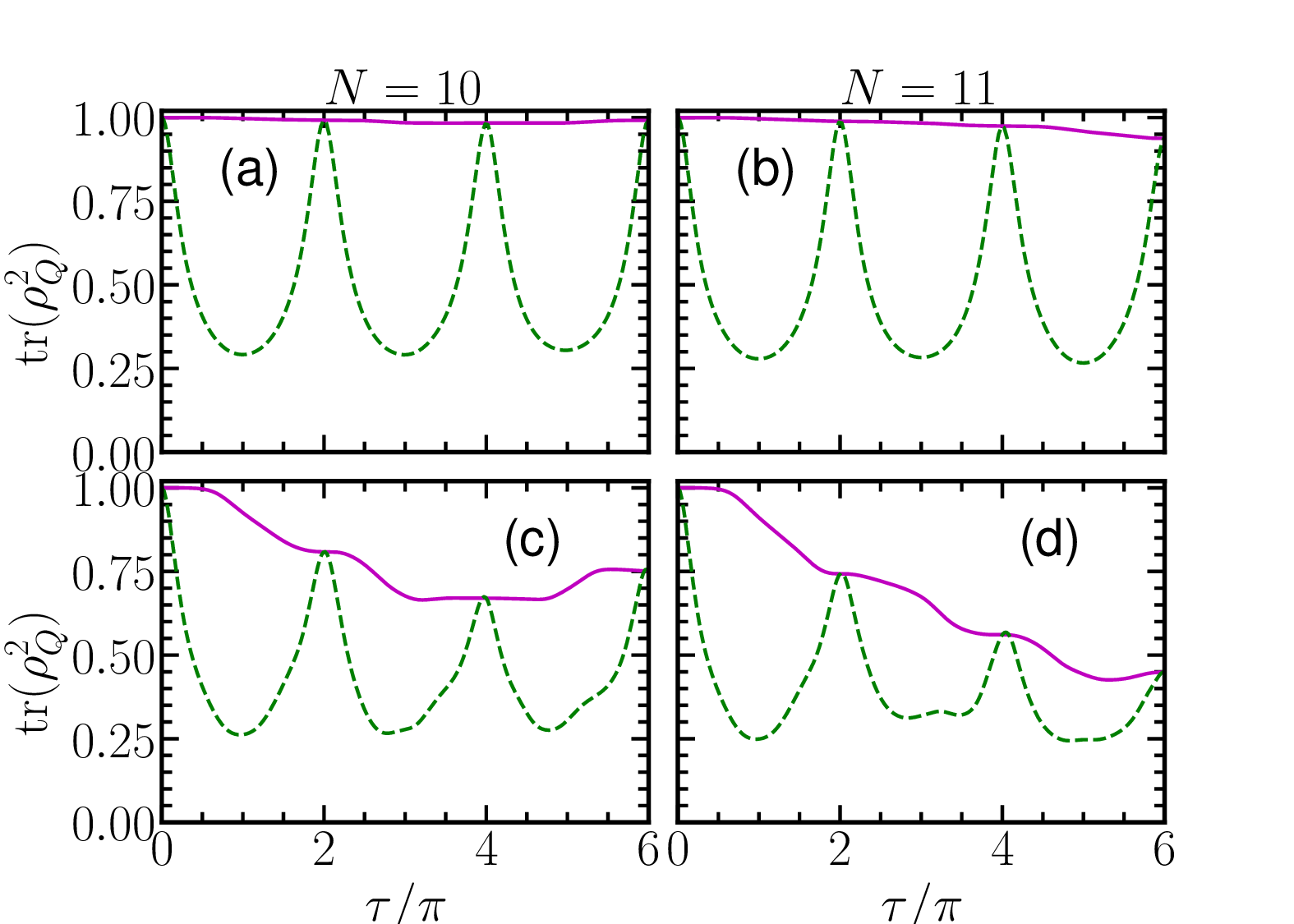}
		\caption{Evolution of purity for $N= 10$ and $11$ spin qubits. The solid line is the purity calculated in the rotated frame of $\lvert\psi'(\tau)\rangle$. The dashed line is the purity calculated in the non-rotated frame $\lvert\psi(\tau)\rangle$. The parameters for the figure are qubit-boson interaction $g = 1/2$ and average boson number amplitude $\alpha = \sqrt{2}$. For the first row, the free oscillation parameter $\epsilon= 0.01$ while in the second row $\epsilon = 0.05$
			}
		\label{fig6}
		\end{figure}

	To understand the effect of free evolution of the qubits on the state preparation, we move into the interaction picture where the state $\lvert \psi'(\tau)\rangle = \hat{U}^{-1}\lvert \psi(\tau)\rangle$ evolves as 
	\begin{equation}
		\label{eq:InteractionPictureDynamics}
		i\frac{d\lvert \psi'\rangle}{d\tau} = H'(\tau)\lvert \psi'\rangle,
	\end{equation}
	and the interaction picture Hamiltonian is $H'(\tau) = \epsilon\hat{U}^{-1} J_z\hat{U} $. The purity $\mathrm{tr}(\rho^2_\mathrm{A})$ of the resulting state $\lvert \psi'(\tau)\rangle$ is shown in Fig.~\ref{fig6}. 
	
	The purity of the qubit states decreases with time, showing some oscillations at small values of free evolution parameter $\epsilon = 0.01$. For an odd number of qubits shown in Fig.~\ref{fig6}(b), the oscillations are barely noticeable. Instead, a slow decrease in purity with a weak free evolution parameter is observed. On the other hand, the purity of the even number of qubits shows the oscillations that decreased more slowly compared to the odd number of qubits (Fig.~\ref{fig6}(a)).

	In the unrotated laboratory frame, the purity of the state puts a cap on the purity attainable in the rotated frame. However, the oscillations and slow decrease in the even spin number case help the fidelity recover even at the long times of $\tau= 6\pi$. Nevertheless, for a significant value of the free evolution parameter $\epsilon = 0.05$ as shown in Figs.~\ref{fig6} (c) and (d),  the purity decreases rapidly. Both the purity of even and odd number of qubits show oscillations. But the purity of the even number of spins decreases slowly compared to the odd number of qubits. The purity in both even and odd numbers of qubits oscillates but never reaches unity. 

    
	The inability of the purity to reach unity shows that the free evolution of the qubits produce a entangled state $\lvert\psi'(\tau)\rangle$ of the qubit and bosons at all times. Thus, there are no times when the quantum states of qubits and bosons are disentangled, except at $\tau =0$. Hence, at times of interest $\tau_{\mathrm{Ent}}$, the state of the qubits are no longer pure and disentangled, and the boson statistics from the unitary operator $\hat{U}$  makes some contributions to the state $\lvert\psi(\tau)\rangle$. Any trace of the bosonic state leaves its statistics on the quantum state of the qubits. Thus, the resulting purity of the qubits in the final state $\lvert\psi(\tau)\rangle$ at $\tau_{\mathrm{Ent}}$ can exceed that of the state $\lvert\psi'(\tau)\rangle$, especially with increases in the strength of free oscillation at long interaction time. This contrasts with the results shown in Fig.~\ref{fig2}, where the spin and bosonic states disentangle at times $\tau_d$.

\begin{figure}[t]
		\includegraphics[width=\columnwidth]{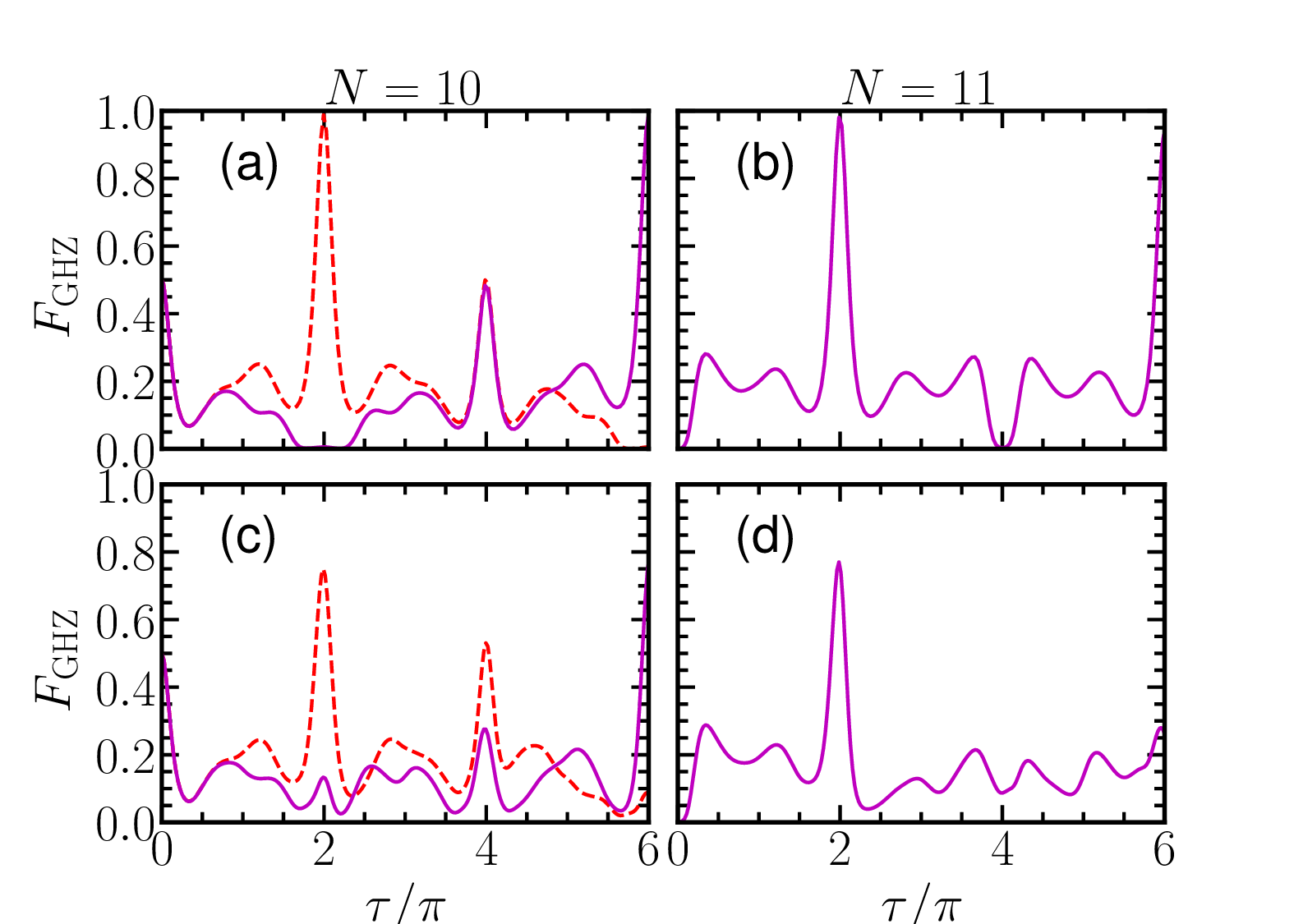}
		\caption{Evolution of the fidelity (\ref{eq:GHZFidelity}) of the GHZ state for $N = 10$ and $11$ qubits. The parameters for the figure are qubit-boson interaction $g = 1/2$ and average boson number amplitude $\alpha = \sqrt{2}$. For the first row, the free oscillation parameter $\epsilon= 0.01$ while in the second row $\epsilon = 0.05$
		}
		\label{fig7}
	\end{figure}

	We further investigate the effect of free oscillations on preparing the Schr\"odinger cat state as shown in Fig.~\ref{fig7}. From the discussion above, the free oscillations prepare a mixed state of the qubits as an input to the unitary $\hat{U}$~(\ref{eq:DisentangledExponentialOperators}), which would impact the preparation of the quantum correlated states. For instance, using the state $\lvert\psi'(\tau)\rangle$ as an input to $\hat{U}$~(\ref{eq:DisentangledExponentialOperators}), $\lvert\psi(\tau)\rangle = \hat{U} \lvert\psi'(\tau)\rangle$, which reduces the fidelity (\ref{eq:GHZFidelity}) of the GHZ states that emerged at times $\tau_{\mathrm{Ent}}$~(\ref{eq:EngineeredEntanglementTime}), see Fig.~\ref{fig7}. This is due to the statistics of the photon state contained in the state $\lvert\psi'(\tau)\rangle$ which contributes to the state preparation. As shown in Fig.~\ref{fig7}(a) and (b), where the free oscillation parameter is weak $\epsilon = 0.01$, the fidelity of the state prepared is nearly unity. However, the impact of using the state $\lvert \psi'(\tau)\rangle$  worsens the fidelity with a significant increase in free oscillation parameter $\epsilon$ as shown in Figs.~\ref{fig7}(c) and~(d).

	More generally, the effects observed with free evolution come from preparing a state which may contain some entanglement between qubits and bosons. Hence, tracing out the bosonic state leaves their statistics on the qubit state. It is these statistics that further corrupt the states of the qubits. It contrasts with the case where free evolution is absent. In such a situation, as demonstrated in Secs.~\ref{sec:EntangledStates} and~\ref{sec:EntangledStatePrep}, the bosonic state does not contribute to the statistics of the qubits at the disentangled times $\tau_d$~(\ref{eq:DisentangledTime}) of which the entangling time $\tau_{\mathrm{Ent}}$~(\ref{eq:EngineeredEntanglementTime}) is of particular interest.

	\section{Experimental Realization\label{sec:Experiment}}
	Recent experiments~\cite{lv2018,koch2023} have demonstrated ultrastrong coupling in various qubit systems, each exhibiting exceptional control and tunability over the qubit frequency $\epsilon$ and qubit-boson interaction strength $g$. For instance, in Ref.~\cite{koch2023}, the dimensionless qubit frequencies, $\epsilon$, were 0, 2.3, and  3.7, while in Ref.~\cite{lv2018}, they were $0$ and  $2$. The studies showed that qubit dynamics with zero qubit frequency, $\epsilon = 0$, have prolonged complete coherence over a longer time than those for which it is greater than zero, in line with our discussions in Secs.~\ref{sec:EntangledStates}, \ref{sec:EntangledStatePrep}, and \ref{sec:FreeEvolution}. Thus, tuning out the qubit frequency makes it inconsequential in the state preparation. Hence, we turn our attention to the qubit-boson interaction strength $g$.
	
	The feasibility of our proposal largely depends on achieving a sufficiently large $g$, as this parameter directly influences the gate operation times. It has been demonstrated that the dimensionless coupling parameter $g$ has reached values as large as $6.5$~\cite{koch2023}. Such coupling strength is quite high compared to decoherence times in those experiments and would allow quantum gate operations to be completed before any decoherence occurs. In our protocol, the gate time is realized at  $\tau = \tau_d$, (\ref{eq:DisentangledTime}). However, the correlations observed at this time would depend on $g$, (\ref{eq:InteractionStrength}). Using dimensionless coupling strength (\ref{eq:InteractionStrength}), and the dimensionless time $\tau = \omega t$, we write the gate time (\ref{eq:DisentangledTime}) in dimensional units as 
	\begin{equation}
		\label{eq:GateTime}
		t_\mathrm{gate} = \frac{\pi \sqrt{2n \beta}}{G}, 
	\end{equation} 
	where $G$ is the qubit-oscillator coupling strength defined in (\ref{eq:RabiHamiltonian}), and the fastest gate time is realized for $n=1$. Thus, we believe that our proposal is within the reach of recent experiments that have achieved large values of qubit-oscillator strength $G$.

	\section{Summary And Conclusions\label{sec:Summary}}
	We analyzed a method for implementing a deterministic multi-qubit entangling gate in the ultrastrong coupling regime, where qubits interact with a shared bosonic mode. This gate expands the toolkit for quantum computing with neutral qubits. At specific times, we showed that the entangling gate~(\ref{eq:DisentangledExponentialOperators}) can be expressed as a product of two unitaries~(\ref{eq:AtomStateAtSpecialPoints}): one that entangles the qubits by acting solely on their state, and another that induces a phase shift by acting on the bosonic state.

    We demonstrated that the tunability of the qubit-boson interaction enables the preparation of diverse correlated states. We showed that for a given number of qubits, the GHZ and squeezed states are prepared by tuning the qubit-boson interaction strength. We demonstrated that achieving a maximally correlated state requires a fast gate time given by (\ref{eq:GateTime}), and is inversely proportional to the coupling strength. For multiple qubit ensembles, such as two ensembles of qubits, the gate produces more distillable entanglement at the disentangling time when the ensemble-boson interaction strengths are different. When the interaction strengths are identical, the amount of distillable entanglement decreases, the negativity approaches a value of one-half, and the quantum state of the ensembles becomes a macroscopic Schrödinger cat state. In particular, the entangling unitary operator for the qubits implies a $ (J^x)^2 $-type interaction, which is well known for generating squeezed ~\cite{kitagawa1993} and entangled states~\cite{byrnes2013}. It also plays a key role in quantum top models employed to study nonlinear dynamics~\cite{dariano1992,chaudhury2009,mondal2020}. Consequently, the tunable coupling strength in the ultrastrong coupling regime of many qubits could facilitate quantum simulations and investigations of exotic phenomena~\cite{simon2011} in quantum nonlinear dynamics, similar to applications in ion trap systems~\cite{sorensen1999,sorensen2000,garcia-ripoll2003}.
    
 	We showed that our proposed gate is robust against weak free qubit oscillations and operates efficiently, with gate times inversely proportional to the coupling strength. This makes the method practical for current ultrastrong coupling experiments, where tuning the free oscillation strength over a broad range presents no significant challenge. Importantly, our approach provides a pathway for achieving many-qubit entanglement in various models governed by the spin-boson Hamiltonian.

\acknowledgments

This work is supported by the National Natural Science Foundation of China (62071301); NYU-ECNU Institute of Physics at NYU Shanghai; the Joint Physics Research Institute Challenge Grant; the Science and Technology Commission of Shanghai Municipality (19XD1423000,22ZR1444600); the NYU Shanghai Boost Fund; the China Foreign Experts Program (G2021013002L); the NYU Shanghai Major-Grants Seed Fund; Tamkeen under the NYU Abu Dhabi Research Institute grant CG008; and the SMEC Scientific Research Innovation Project (2023ZKZD55).

	\appendix
	\section{Displacement of photon operators\label{sec:DisplacementOperator}}
	More generally consider the following operator 
	\begin{equation}
		\label{eq:UnitaryGen}
		\hat{U} = e^{-\beta\hat{a}^\dagger\hat{a} -\eta (\hat{a}^\dagger + \hat{a})}.
	\end{equation}
This displaces the operators $\hat{a}^\dagger$ and $\hat{a}$ as follows
	\begin{align}
		\label{eq:UnitaryOnAnnihilation}
		\hat{U}^{-1}\hat{a}\hat{U} & = \hat{a}e^{-\beta} -\frac{\eta}{\beta}\left(1 - e^{-\beta}\right),\\
		\label{eq:UnitaryOnCreation}
		\hat{U}^{-1}\hat{a}^\dagger\hat{U} & = \hat{a}^\dagger e^{\beta} +\frac{\eta}{\beta}\left(e^{\beta} - 1\right).
	\end{align}

	\section{Disentangling exponential operators\label{sec:DisentanglingExponentialOperators}}
	Consider an exponential operator function $F$ of the form
	\begin{equation}
		\label{eq:ExponentialOperatorFiunction}
		F = e^{-i\tau\left[\hat{a}^\dagger\hat{a}+ gJ^x(\hat{a} + \hat{a}^\dagger) \right]}.
	\end{equation}
	The exponential operator is diagonal in qubit operators, but not in the photon operators. We then seek a product of exponential operators in the photon operators. Note that the operators satisfy the following commutation relations
	\begin{equation}
		\label{eq:PhotonCommutationRelation}
		\begin{split}
			[J^x\hat{a},J^x\hat{a}^\dagger] & = (J^x)^2\\
			[\hat{a}^\dagger\hat{a},J^x\hat{a}] &= -J^x\hat{a}\\
			[\hat{a}^\dagger\hat{a},J^x\hat{a}^\dagger] & = J^x\hat{a}^\dagger
		\end{split}
	\end{equation} 
	Hence, any equivalent formulation of $F$ arranged in normal ordering must contain these operators and a constant operator of the form
	\begin{equation}
		\label{eq:EquivalentExponentialOperator}
		F = e^{P(\tau)J^x \hat{a}^\dagger} e^{Q(\tau)\hat{a}^\dagger\hat{a}} e^{R(\tau)J^x \hat{a}} e^{S(\tau)(J^x)^2 }.
	\end{equation}
	Differentiating (\ref{eq:ExponentialOperatorFiunction}) with respect to $\tau$ gives 
	\begin{equation}
		\label{eq:DifferentialExpeonetialOperator}
		\frac{dF}{d \tau} = -i\left[\hat{a}^\dagger\hat{a} + gJ^x\left(\hat{a}^\dagger + \hat{a}\right) \right] F .
	\end{equation}
	Similarly differentiating the equivalent formulation (\ref{eq:EquivalentExponentialOperator}) with respect to $\tau$ gives
	\begin{equation}
		\label{eq:DifferentialEquivalentFormulation}
		\begin{split}
			\frac{dF}{d \tau} = \bigg(\dot{P} J^x \hat{a}^\dagger+ e^{PJ^x\hat{a}^\dagger}\dot{Q}\hat{a}^\dagger\hat{a}e^{-PJ^x\hat{a}^\dagger}\\
			+ e^{PJ^x\hat{a}^\dagger} e^{Q\hat{a}^\dagger\hat{a}} \dot{R}J^x\hat{a} e^{-Q\hat{a}^\dagger\hat{a}} e^{-PJ^x\hat{a}^\dagger} + \dot{S}(J^x)^2\bigg)F,
		\end{split}
	\end{equation}
	where the dots on $P$, $Q$, $R$ and $S$ represent their derivative with the dimensionless time $\tau$. 
	
	Applying the Baker-Campbell-Hausdorff relation to (\ref{eq:DifferentialEquivalentFormulation})~\cite{louisell1990} and equating coefficients of operators in (\ref{eq:DifferentialExpeonetialOperator}) and (\ref{eq:DifferentialEquivalentFormulation}) gives the following coupled differential equations 
	\begin{equation}
		\label{eq:CoupledDifferentialEquations}
		\begin{split}
			\dot{P} - P\dot{Q} & = -ig, \\
			\dot{Q} & = -i,\\
			\dot{R} e^{-Q}& = -ig,\\
			\dot{S} - P\dot{R}e^{-Q} & = 0. 
		\end{split}
	\end{equation}
	Solving (\ref{eq:CoupledDifferentialEquations}) subject to the initial condition $P(0) = Q(0) = R(0) = S(0) = 0$ gives the following
	\begin{align}
		\label{eq:ppSolution}
		P(\tau) & = -g(1 -e^{-i\tau}),\\
		\label{eq:qqSolution}
		Q(\tau) & =  -i\tau,\\
		\label{eq:rrSolution}
		R(\tau) & = -g(1 -e^{-i\tau}),\\
		\label{eq:ssSolution}
		S(\tau) & = ig^2\left(\tau + i\left(1 - e^{-i\tau}\right)\right) .
	\end{align}	
    Using $P = ip$, $Q=iq$, $R = ir$, and $S=is$ we obtain the equations used in the main text.
	
	Note that for anti-normal ordering of the exponential operators
	\begin{equation}
		\label{eq:AntinormalOrdering}
		F = e^{R(\tau)J^x \hat{a}}e^{Q(\tau)\hat{a}^\dagger\hat{a}}e^{P(\tau)J^x\hat{a}^\dagger} e^{S(\tau)(J^x)^2},
	\end{equation}
	and following a similar procedure as above, one obtains
	\begin{align}
		\label{eq:AppSolution}
		P(\tau) & = g(1 -e^{i\tau}),\\
		\label{eq:AqqSolution}
		Q(\tau) & =  -i\tau,\\
		\label{eq:ArrSolution}
		R(\tau) & = g(1 -e^{i\tau}),\\
		\label{eq:AssSolution}
		S(\tau) & = ig^2\left(\tau + i\left(e^{i\tau} - 1\right)\right) .
	\end{align}	

	\section{Ensembles interacting with a common bosonic mode\label{sec:EnsembleCase}}
    Consider an ensemble of qubits, each interacting with a common bosonic mode. The \emph{k}\textsuperscript{th} ensemble contains $N_k$ qubits and has a qubit-boson interaction strength $G_k$. The total spin $J_k$ of each ensemble is $J_k=N_k/2$ and the spin of each spin ensemble commute $[J^{\alpha}_n,J^{\beta}_k] = \delta_{n,k}\epsilon_{\alpha\beta\gamma}J^\gamma_k$, where $\epsilon_{\alpha\beta\gamma}$ is the Levi-Civita anti-symmetric tensor and $\alpha,\,\beta,\,\gamma \in \{x,y,z\}$. Thus, the Hamiltonian describing the interaction of \emph{M} ensembles with a common bosonic mode is
    \begin{equation}
        \label{eq:EnsembleHamiltonian}
        H = \hbar\sum_{k=1}^{M}\omega_{0,k} J^{z}_{k} + \hbar\omega\hat{a}^\dagger\hat{a} + \hbar\sum_{k=1}^{M} G_{k}J^x_k(\hat{a}^\dagger + \hat{a}),
    \end{equation}
    where $\omega_{0,k}$ is the qubit frequency of the \emph{k}\textsuperscript{th} ensemble.

    Following the discussions of Sec.~\ref{sec:CollectiveSpins}, we write the dimensionless Hamiltonian $H_0/(\hbar\omega)$ for the many ensemble case as
    \begin{equation}
        \label{eq:EnsembleH0Hamiltonian}
        H_0 =  \hat{a}^\dagger\hat{a} + \sum_{k=1}^{M} g_{k}J^x_k(\hat{a}^\dagger + \hat{a}),
    \end{equation}
    where $g_k = G_k/\omega$. The unitary operator that evolves the initial state of the \emph{M} ensembles to another state at time $\tau$ is 
    \begin{equation}
        \label{eq:ManyEnsembleUnitary}
        \hat{U} = e^{-i\tau\left[ \hat{a}^\dagger\hat{a} + \sum_{k=1}^{M} g_{k}J^x_k(\hat{a}^\dagger + \hat{a}) \right]}.
    \end{equation}
    Using the methods of Appendix~\ref{sec:DisentanglingExponentialOperators}, we write (\ref{eq:ManyEnsembleUnitary}) in ordered form as (\ref{eq:EquivalentExponentialOperator}), where $J^x = \sum_{k=1}^{M} g_{k}J^x_k$.

	\bibliographystyle{apsrev4-2}

	
	%

 \end{document}